\documentclass[prl,preprint,showpacs]{revtex4}
\usepackage{graphicx}
\begin{document}

\title{Interface driven reentrant superconductivity in HoNi$_5$-NbN-HoNi$_5$ nanostructures}

\author{Gyanendra Singh$^1$, P. C. Joshi$^1$, Z. Hossain$^1$ and R. C. Budhani$^{1,2}$$^\ast$}
\affiliation{$^1$Condensed Matter - Low Dimensional Systems
Laboratory, Department of Physics, Indian Institute of Technology
Kanpur, Kanpur - 208016, India\\
$^2$National Physical Laboratory, Council of Scientific and Industrial Research, New Delhi-110012, India}
\email{rcb@iitk.ac.in, rcb@nplindia.org}
\begin{abstract}

Superconductivity (S) and ferromagnetism (F) are probed through transport and magnetization measurements in nanometer scale HoNi$_5$-NbN (F-S) bilayers and HoNi$_5$-NbN-HoNi$_5$ (F-S-F) trilayers. The choice of materials has been made on the basis of their comparable ordering temperatures and strong magnetic anisotropy in HoNi$_5$. We observe the normal state reentrant behavior in resistance vs. temperature plots of the F-S-F structures just below the superconducting transition in the limited range of HoNi$_5$ layer thickness d$_{HN}$ (20 nm $<$ d$_{HN}$ $<$ 80 nm) when d$_{NbN}$ is fixed at $\simeq$ 10 nm. The reentrance is quenched by increasing the out-of-plane (H$_{\perp}$) magnetic field and transport current where as in-plane (H$_{\parallel}$) field of $\leq$ 1500 Oe has no effect on the reentrance. The thermally activated flux flow characteristics of the S, F-S and F-S-F layers reveal a transition from collective pinning to single vortex pinning as we place F layers on both sides of the S film. The origin of the reentrant behavior seen here in the range of 0.74 $\leq$ T$_{Curie}$/T$_C$ $\leq$ 0.92 is attribute to a delicate balance between the magnetic exchange energy and the condensation energy in the interfacial regions of the trilayer.

\end{abstract}
\maketitle
\section{introduction}
The antagonistic order parameters of superconductor (S) and ferromagnet (F) lead to several fascinating effects in the transport and magnetic properties of thin film S-F heterostructures \cite{Buzdin,Lyuksyutov}. It has been noticed that there is a large suppression of the superconductivity at the S-F interface due to the strong pair breaking effect of the ferromagnet via spin-flip scattering and/or spin rotation. The ferromagnetic layer is also affected by the presence of the superconductor as the Cooper pairs entering the F region acquire a center of mass momentum due to the exchange field of the F. This adds an oscillating term to the Cooper pair wave function inside the ferromagnetic region \cite{Demler}. The experimental studies on S-F bilayers \cite{Muhge,Cirillo,Zdravkov,Kehrle}, trilayers \cite{Garifullin,Tikhonov,Obi} multilayers \cite{Verbanck,Jiang,Goryunov,Mercaldo} and S-F-S junctions \cite{Kontos,Ryazanov,Oboznov,Robinson,Khaire,Witt} show effects such as the direct and inverse proximity effect, reentrant superconductivity and critical temperature oscillations.

In addition, the phenomenon of long range triplet pairing is possible when the coupled ferromagnetic layers have inhomogeneities in magnetization, which can be due to domain walls, spiral magnetism and spin scattering at the interface \cite{Witt,Khaire,Bergeret}. In the case of a F-S-F trilayer, for certain angles between the magnetization vectors of the F layers, a triplet pairing can be induced in the F layers \cite{Leksin,Zhu}. As a result, the Cooper pair can survive in ferromagnetic region up to length scale of normal metal coherence length ($\xi$$_N$).

In most of the S-F hybrids studied till date, the F order sets in at a temperature (T$_{Curie}$) higher than the transition temperature (T$_C$) of the superconductor and thus the nucleation of the superconducting state is subjected to a robust exchange field of the ferromagnet. An equally important and perhaps much more illuminating option is to have a ferromagnet whose T$_{Curie}$ is lower than the T$_{C}$ of the superconductor. If the ferromagnet is sufficiently thin, one may see suppression of T$_{Curie}$ due  to induced superconducting order in the F. To the best of our knowledge, there is no direct evidence of this, while the converse has been well demonstrated.

The heterostructures of NbN and HoNi$_5$ offer an ideal system to study suppression of F order due to superconductivity because NbN becomes superconducting at $\sim$16 K and the bulk ferromagnetic ordering temperature (T$_{Curie}$) of HoNi$_5$ $\sim$5 K. In our previous article on NbN/HoNi$_5$ bilayers \cite{Joshi}, we have reported a spectacular flux flow induced peak effect in the magnetoresistance R(H) of the system at temperatures below which HoNi$_5$ is ferromagnetic. This observation was attributed to a spin reorientation transition induced by the in-plane magnetic field. These bilayer samples, however, do not show any direct signatures of the competition between the F and S order parameters. In contrast, a trilayer film of HoNi$_5$/NbN/HoNi$_5$ displays dramatic effect of the antagonism in R(T) and M(T) measurements as the system is cooled through the superconducting and magnetic transition temperatures. The most striking signature of the competition is a reentrant superconducting transition in the R(T) data of the trilayers when the NbN films are made sufficiently thin such that the perturbation from the F layer is strong enough to suppress superconductivity over a limited range of temperature below the T$_C$ of NbN. We have established a critical range of T$_{Curie}$/T$_C$, where T$_{Curie}$ and T$_C$ respectively are magnetic and superconducting transition temperature of the trilayer in which the reentrance is seen. The reentrant behavior is also seen prominently in the M(T) data. We draw a comparison between these results and a recent prediction of reentrant superconductivity (RES) in F-S layers where F layer has inhomogeneous magnetization \cite{Chien}. Since HoNi$_5$ is a highly anisotropic ferromagnet and the films used here are polycrystalline with evidence of inhomogeneous magnetization, it may allow a strong superconducting proximity effect at lower temperatures. However, we can not rule out the possibility that Ho 4f moments at both interfaces perturb the superconductivity of the thin NbN layer, leading to reentrance to the normal state. To the best of our knowledge we are presenting first observations of reentrant superconductivity in S-F heterostructures of T$_{Curie}$ $<$ T$_C$, while it has been observed prominently in bulk superconductors such as ErRh$_4$B$_4$, HoMo$_6$S$_8$ and HoNi$_2$B$_2$C \cite{Fertig,Crabtree,Eisaki,Rathnayaka,Gupta}.

\section{experimental}
Thin films of NbN, HoNi$_5$ and their heterostructures were deposited on (100) cut MgO substrates using pulsed laser deposition technique. For HoNi$_5$, a stoichiometric polycrystalline target was ablated in 0.1 mbar neon environment and then a NbN film was deposited in 0.1 mbar N$_2$ pressure on top of the HoNi$_5$ layer. For the deposition of the top HoNi$_5$ layer, the nitrogen was flushed out and the growth was carried out at 0.1 mbar of neon. The growth temperature for all three layers was 200$^0$ C. The thickness d$_{HN}$ of the bottom and top HoNi$_5$ layers was kept the same but varied from 20 to 80 nm in different samples and the NbN thickness (d$_{NbN}$) was changed from 10 to 30 nm for given d$_{HN}$. Further details of growth condition can be found in previous reports \cite{Gyanendra,Senapati}. For the measurements of electrical transport, the samples were patterned into 15 $\mu$m wide lines by using tungsten shadow mask and Ar$^+$ ion milling technique. A schematic diagram of sample geometry is shown in the upper inset of Fig. 1(a).

\section{Results and discussion}
We first present the results of x-ray reflectivity measurements on HoNi$_5$-NbN bilayers of d$_{HN}$ = 50 nm to quantify the extent of interface roughness. These data are shown in Fig. 1(a).  A best fitting of a genetic algorithm (solid line) yields a roughness of 0.7 and 1.3 nm for NbN/HoNi$_5$ bilayer with 30 and 10 nm thick NbN respectively. The superconducting transitions as seen in resistivity measurements on HoNi$_5$-NbN-HoNi$_5$ trilayers of NbN thickness 10 and 30 nm are displayed in the lower inset of Fig. 1(a). The trilayer with 30 nm NbN shows a sharp transition with critical temperature (T$_C$) of $\sim$11 K. On reducing the thickness to $\sim$10 nm, the onset of T$_C$ shifts down to $\approx$ 6 K. This drop, however, is truncated by a valley and then upturn of R(T) on reducing the temperature below $\approx$ 5.5 K. The resistance goes through a peak at $\approx$ 5 K followed by a sharp drop at lower temperatures. This behavior is similar to the reentrant phenomenon seen in bulk samples of HoNi$_2$B$_2$C, ErRh$_4$B$_4$ and HoMo$_6$S$_8$ \cite{Fertig,Crabtree,Eisaki,Rathnayaka,Gupta}. A magnified view of the reentrant nature of the R(T) of the trilayer with 10 nm thick NbN is shown in panel (b) of Fig. 1. The R(T) is separated into three regions: a normal metallic behavior for T $>$ 6 K, reentrant superconductivity (red bar) between 5 K $<$ T $<$ 6 K and zero resistance state for T $<$ 5 K. No such reentrant behavior was seen in the trilayers with 30 nm thick NbN. In the inset of Fig. 1(a), we also display the R(T) data for a bare 10 nm thick NbN film. From a comparison of curve A and B, it is clear that the T$_C$ is suppressed drastically in the trilayer for a fixed NbN layer thickness.

At this stage it is important to recall the magnetic properties of HoNi$_5$, which shows bulk ferromagnetic ordering at $\sim$5 K. The magnetism in this compound is highly anisotropic due to 4f electrons of Holmium. The easy axis of magnetization lies in the ab-plane of the hexagonal unit cell, but the magnetization for field along the c-axis does not saturate at field as high as 15 T \cite{Gignoux}. The HoNi$_5$ thin films grown on (100) MgO investigated by our group \cite{Gyanendra} are polycrystalline in nature with grain size in the range of $\approx$ 10 nm. To establish magnetism and superconductivity in HoNi$_5$/NbN/HoNi$_5$ trilayer with 10 nm thick NbN, we have measured temperature dependent magnetization M(T) after zero field cooling the sample and then applying 100 Oe field along the out-of-plane direction. The trilayer used in magnetization measurements was grown along with the sample used for transport measurements in the same run. The M(T) shown in Fig. 1(c) rises sharply below 5 K, goes through a peak and then drop to a negative value, indicating strong diamagnetism. A magnified view of M(T) in the temperature range of 3 to 7 K is shown in the inset of the figure along with the M(T) of a bare HoNi$_5$ film on MgO \cite{Gyanendra}. The T$_{Curie}$ extracted from the behavior M in the crtical regime $\sim$ ${M_0}{(1 - T/{T_{Curie}})^\beta }$ is $\approx$ 5.5 K and $\beta$ $\approx$ 0.52. As we see in Fig. 1(b) the onset of superconductivity is at $\approx$ 6 K. However, before a fully diamagnetic state could develop, the fluctuating Cooper pair density experiences the pair breaking field of the HoNi$_5$ and hence there is no substantial change in the rising part of the M(T) curve. However, at still lower temperature the condensate becomes robust and a competition between diamagnetism and ferromagnetism leads to a peak in M(T) at $\approx$ 3.5 K. On further lowering the temperature, the diamagnetism of superconducting NbN clearly dominates the ferromagnetic response of the system. The appearance of the reentrance in trilayers also depends on the thickness (d$_{HN}$) of each HoNi$_5$ layer. As seen in Fig. 1(d), when d$_{HN}$ is sufficiently thick ($\simeq$ 80 nm), the T$_C$ of NbN is greatly suppressed with no evidence of reentrance. Also when d$_{HN}$ is small ($\simeq$ 20 nm), the T$_C$ of NbN remains robust with no sign of entry into the normal state on cooling below T$_C$. In Fig. 1(e) we show the variation of T$_{Curie}$ of trilayer as a function of the total thickness d$_F$ (=d$_{F1}$ + d$_{F2}$) of HoNi$_5$ layers. The T$_{Curie}$ drop by $\simeq$ 1.4 K on lowering d$_F$ from 160 nm to 40 nm. From these data a very interesting correlation evolves between the T$_{Curie}$/T$_C$ of the F-S-F and the observation of the reentrance. The later is seen when 0.74 $\leq$ T$_{Curie}$/T$_C$ $\leq$ 0.92 (see inset of Fig. 1(e)).

In order to get further insight of the reentrant behavior of superconductivity, we have measured the resistivity of the sample in the transition region in a magnetic field applied along the out-of-plane (H$_{\perp}$) and in-plane (H$_{\parallel}$) directions (perpendicular to bar) of the trilayer. The results of in-plane field measurement are shown in Fig. 2(a). The superconducting transition in the H$_{\perp}$ configuration is much more sensitive to the field; it becomes significantly broader as we increase the field from 0 to 1400 Oe. An enlarge view of the behavior of minimum and peak in R(T) which characterize the reentrant transition is shown in the inset of Fig. 2(b) for several values of H$_{\perp}$. The changes in the minimum have been quantified in terms of  $\Delta R = {R_{p}} - {R_{\min }}$ and T$_{min}$(H$_{\perp}$), and are plotted in the lower inset of Fig. 2(b). The $\bigtriangleup$R and T$_{min}$(H$_{\perp}$) drop monotonically with the applied field. The H$_{\perp}$ also leads to a large number of vortices in the system. The thermally activated flux flow (TAFF) can lead to the broadening of the resistive transition in the tail region of the curve. We have analyzed this behavior in the framework of TAFF model \cite{Anderson} in the subsequent section.

The reentrant behavior shows a strong dependence on the current used for the resistivity measurements. In Fig. 3(a) we show the R(T) data in zero field for current varying from 1 $\mu$A to 2 mA (current density j from 6.6 x 10$^2$ A/cm$^2$ to 133.3 x 10$^4$ A/cm$^2$). The changes in the R(T) with current in the temperature regime where the two order parameters compete strongly, is quite different from the behavior seen under H$_{\perp}$ field (Fig. 2(b)). Here at the lowest current density, there is very little evidence of a minimum in the R(T) below the onset of superconductivity. We only see a shoulder below which the resistance drops precipitously to zero. As the current density is increased from 6.6 x 10$^2$ A/cm$^2$ to 6.6 x 10$^4$ A/cm$^2$, the minimum in the R(T) becomes pronounced and the T$_p$ shows a small shifts to lower temperatures (see Fig. 3(b)). A further increase in j to 33.3 x 10$^4$ A/cm$^2$ leads to saturation of $\Delta R$. However, for j $\geq$ 33.3 x 10$^4$ A/cm$^2$ a flattening of the minimum and shift of the T$_P$ to lower values is observed (not shown here). The R(T) in the tail portion of the transition (T $<$ T$_P$) becomes progressively broader on increasing the j from 6.6 x 10$^2$ A/cm$^2$. While the reasons for this strong current dependence is not understood fully, it is certainly not a heating effect. To further clarify the issue of sample heating, we have measured current vs. voltage (I-V) curves at different temperatures across the superconducting transition in forward and reverse directions. We observe no hysteresis in IVs which could have resulted from a thermal lag between the temperature seen by the sensor and the actual temperature of the sample. A similar current dependence of the reentrant behavior has been reported by Rathnayaka \emph{et} \emph{al}. in HoNi$_2$B$_2$C samples \cite{Rathnayaka}. There authors have reported an interesting correlation between the magnitude of the current dependence and the ratio of magnetic and superconducting transition temperature. The effect is seen when  T$_m$/T$_c$ $\geq$ 0.6. The value of T$_m$/T$_c$ in our case is $\simeq$ 0.88. Furthermore, our observation of very little reentrant behavior seen at low currents is also similar to the behavior observed in HoNi$_2$B$_2$C crystal \cite{Rathnayaka}.

In order to understand how the dissipative behavior of thin NbN in the thermally activated flux flow (TAFF) regime is affected by the formation of F-S-F structures, we have compared the R(T) of a plain 10 nm NbN film with that of a F-S and F-S-F structures, having the same S layer thickness (10 nm). These measurements were performed in H$_{\perp}$ varying in the range of 50 to 1500 Oe. Fig. 4 shows the Arrhenius plots of normalized resistance (lnR/R$_n$ vs. 1/T) for all three films. We notice a significant broadening of the resistive transition with field as we go from the single layer to bilayer and then to trilayer. The activation energy for dissipation are calculated using the model used by Hsu and Kapitulnik \cite{Hsu} for two dimensional (2D) superconducting film in which the coherence length $\xi$(T) and penetration depth $\lambda$(T) are comparable with the film thickness. Since, for NbN $\xi$(0) $\sim$ 5 nm and $\lambda$(0) $\sim$ 200 nm, and the NbN layer thickness is only $\simeq$10 nm, the 2D condition is satisfied approximately in this case. Based on the model, the R(T) at the bottom of the superconducting transition should follow the relation $\ln (R/{R_N}) =  - {U_0}(H)/{k_B}T + K(H)$ where U$_0$(H) is zero temperature activation energy and K(H) is coefficient of linear temperature correction. Since the low temperature curve is linear, the U$_0$(H) is obtained by calculating the slope of the lnR/R$_n$ vs. 1/T curve. The results of this analysis are presented in Fig. 5(a). We notice a significant drop in U$_0$(H) as we go from the plain NbN to F-S and then F-S-F. This result can be attributed to the suppression of superconducting order parameter at F-S interface due to the proximity coupling with HoNi$_5$. Such perturbation on the S order parameter is expected to be more in a F-S-F film as both surfaces of the superconducting film see a magnetic layer in this case. This amounts to an effective reduction in the thickness of the film by 2$\xi$$_S$ where $\xi$$_S$ is coherence length in S. Similar behavior is also observed by Senapati \emph{et.} \emph{al.} on high T$_C$ based F-S-F structure \cite{Budhani} where they have shown that U$_0$(H) follow linear relation with thickness of S layer \cite{White}.

We further examine the nature of dissipation in all three films by investigating the slope of the field dependence activation energy U$_0$(H). The inset of Fig. 5(a) shows U$_0$(H) in a log-log plot. The activation energy of the films at low fields follows the power law relation $U(H) \sim {H^{ - \alpha }}$ with $\alpha$ values of $\approx$ 0.45 , 0.33 and 0.2 for the S, F-S and F-S-F thin films respectively. Such variation in the value of the exponent has been observed previously and has been explained on the basis of collective vortex creep for $\alpha$ $\approx$ 1, plastic flux creep for $\alpha$ $\approx$ 0.5 and single vortex pinning for $\alpha$ $\approx$ 0.1 \cite{Palstra,Kucera,Budhani,Brunner}. The observed value of $\alpha$ $\approx$ 0.45 for bare S film suggests the possibility of plastic deformation of flux line lattice in weakly pinning vortex liquid. Similar value is also reported by Fogel \emph{et.} \emph{al.} at low fields in Mo/Si multilayer system \cite{Fogel}. The presence of HoNi$_5$ layer appears to enhance vortex pinning, leading to a single vortex depinning dynamics.

We now address the strong measuring current dependence of the dissipation in the tail region of the superconducting transition in the F-S-F film \cite{Abulafia}. The data in the tail region of the transition are plotted as a function of 1/T in the insert of Fig. 5(b). The behavior of activation energy for a fixed j value is shown in the main panel. The U$_0$(H) remains nearly constant for j $<$ 10 x 10$^4$ A/cm$^2$ but beyond this value a sharp drop in the activation energy is seen. This suggest a transition to the collective creep regime of vortex motion at high current.

We close the discussion by proposing a possible scenario for the observation of reentrance in superconductivity in our magnetic thin film heterostructures. It is important to point out here that our observation of reentrant superconductivity as a function of temperature appears different from the so called reentrant behavior recently reported in Nb-Cu$_{41}$Ni$_{59}$ bilayers as a function of CuNi layer thickness \cite{Zdravkov,Kehrle}. Such observations have been made earlier as well \cite{Muhge,Cirillo}. This arises because the superconducting pair amplitude (PA) is oscillating inside the F, and depending on the thickness of the F layer \cite{Buzdin,Lyuksyutov}, the PA at the interface region of the F can be very large or nearly zero, leading to a higher or lower T$_c$ respectively. A sudden entry into (or out of) the normal state can also occur if the magnetization vectors of the F layers transit from a parallel to antiparallel configuration as a function of temperature. These spin accumulation \cite{Gennes} and Cooper pair averaging effect have been addressed in the past \cite{Fominov}. However, such magnetization reorientation transition is more likely to occur in a magnetic field than by temperature. Recently Wu, Valls and Halterman \cite{Chien} have proposed an interesting possibility of observing a temperature dependent reentrant superconductivity in a F-S film if the F layer has a spiral magnetic order, such as that exists in Holmium. The magnetic structure of HoNi$_5$ is not fully understood, although magnetization measurements show strong anisotropy with $\emph{a}$-b plane of hexagonal unit cell being the easy plane. The HoNi$_5$ films grown on MgO substrate shows inhomogeneities in magnetization due to different orientation of nanostructures, as we have discussed in previous report \cite{Gyanendra}. It is possible that the inhomogeneous magnetization of HoNi$_5$ allows a reentrant behavior in the Wu, Valls and Halterman sense \cite{Chien}. This effect is presumably accentuated by the presence of HoNi$_5$ on both sides of the thin NbN layer.

A more intuitive explanation for the RES can, however, be given in terms of proximity effect in a normal (N) and ferromagnetic metal. The decay of superconducting order parameter in N and F layers is given by the length scale  ${\xi _N} = {\left( {\hbar {D_N}/2\pi {k_B}{T_{CS}}} \right)^{1/2}}$ and $\xi _F^* = {\left( {{{\hbar {D_F}} \mathord{\left/ {\vphantom {{\hbar {D_F}} {2\pi {K_B}{T_{CS}}}}} \right. \kern-\nulldelimiterspace} {2\pi {K_B}{T_{CS}}}}} \right)^{1/2}}$, respectively \cite{Cirillo}. Here $\xi _N$ and $\xi _F^*$ are coherence length, and ${D_N}$ and ${D_F}$ are diffusion coefficients in the normal and ferromagnetic metal respectively. The other relevant length scale in the problem is ${\xi _F} = {\left( {{{\hbar {D_F}} \mathord{\left/ {\vphantom {{\hbar {D_F}} {{E_{ex}}}}} \right. \kern-\nulldelimiterspace} {{E_{ex}}}}} \right)^{1/2}}$, which defines the decay of the S order parameter in the F layer. It is measure of the length scale over which superconductivity is induced in the F layer. The exchange splitting of the conduction band ${E_{ex}}$ is related to the exchange integral \emph{I} and magnetic moment ${\mu _F}$ as ${E_{ex}} = I{\mu _F}$ \cite{Aarts}. Since ${E_{ex}}$ goes to zero at T$_{Curie}$, ${\xi _F}$ diverges on warming the sample toward the magnetic ordering temperature. This divergence is however cut off by $\xi _N$. This is the unique feature of our experiment because T$_{Curie}$ $<$ T$_C$. As long as T$_{Curie}$ $<$ T $<$ T$_C$, the superconductivity in the effective thickness of d$_S$ + 2$\xi$$_N$ where d$_S$ is the thickness of superconducting layer, leads to a drop of resistance. But as soon as T $\simeq$ T$_{Curie}$ this effective thickness starts decreasing not only because ${\xi _F}$ $<$ ${\xi _N}$ but also because of the pair breaking effects of the magnetic layer inside the superconducting film. The consequence of this would be a rise in the resistance. We have estimated the diffusion coefficient ${D_F}$ of HoNi$_5$ from the specific heat data \cite{Sankar,Bechman} and measured resistivity of 40 nm thick HoNi$_5$ film, which is $\approx$ 280 $\mu$$\Omega$cm. Since the elastic mean free path for this value of resistivity is already of the order of interatomic distance, we expect a marginal increase of resistivity of the thinner HoNi$_5$ film due to size effect \cite{Lee}. Band structure calculations of ${E_{ex}}$ for HoNi$_5$ \cite{Malik} suggest a value of 0.5 eV, which yields ${\xi _F}(0)$ $\approx$ 0.3 nm. Since for a strong ferromagnet like Fe, ${\xi _F}$ is only $\approx$ 1 nm \cite{Leksin}, we believe that this calculation overestimates ${E_{ex}}$. The ${\xi _N}$ on the other hand is 2.7 nm. As the correlation length in HoNi$_5$ reduces on going through the magnetic transition, a reentrance into the normal state is expected. However, at still lower temperatures, the order parameter in the S film becomes robust enough to short circuit the HoNi$_5$ films and the resistance would start approaching zero value. These semi quantitative arguments are consistent with the results in Fig. 1.

\section{conclusion}
In summary, we have carried out a detail study of electronic transport in F-S and F-S-F thin film where the F and S order parameters are of comparable strength. In the F-S-F trilayer we see a robust reentrant superconductivity over a critical range of the ratio of magnetic and superconducting transition temperature of the F-S-F trilayer (0.74 $\leq$ T$_{Curie}$/T$_C$ $\leq$ 0.92). The RES seen here is different from the more common reentrant behavior observed as a function of sample geometry. Many features of the RES reported here are similar to those found in bulk magnetic superconductors consisting of ternary and quaternary alloys of rare earths \cite{Fertig,Crabtree}.

\section{acknowledgments}
This research has been supported by a grant from Department of Information Technology (DIT), India. Gyanendra Singh acknowledges financial support from the Council of Scientific and Industrial Research (CSIR), Government of India. RCB acknowledges financial support from J.C. Bose Fellowship of the Department of Science and Technology, India.

\begin{figure}[h]
\begin{center}
\includegraphics [width=5cm]{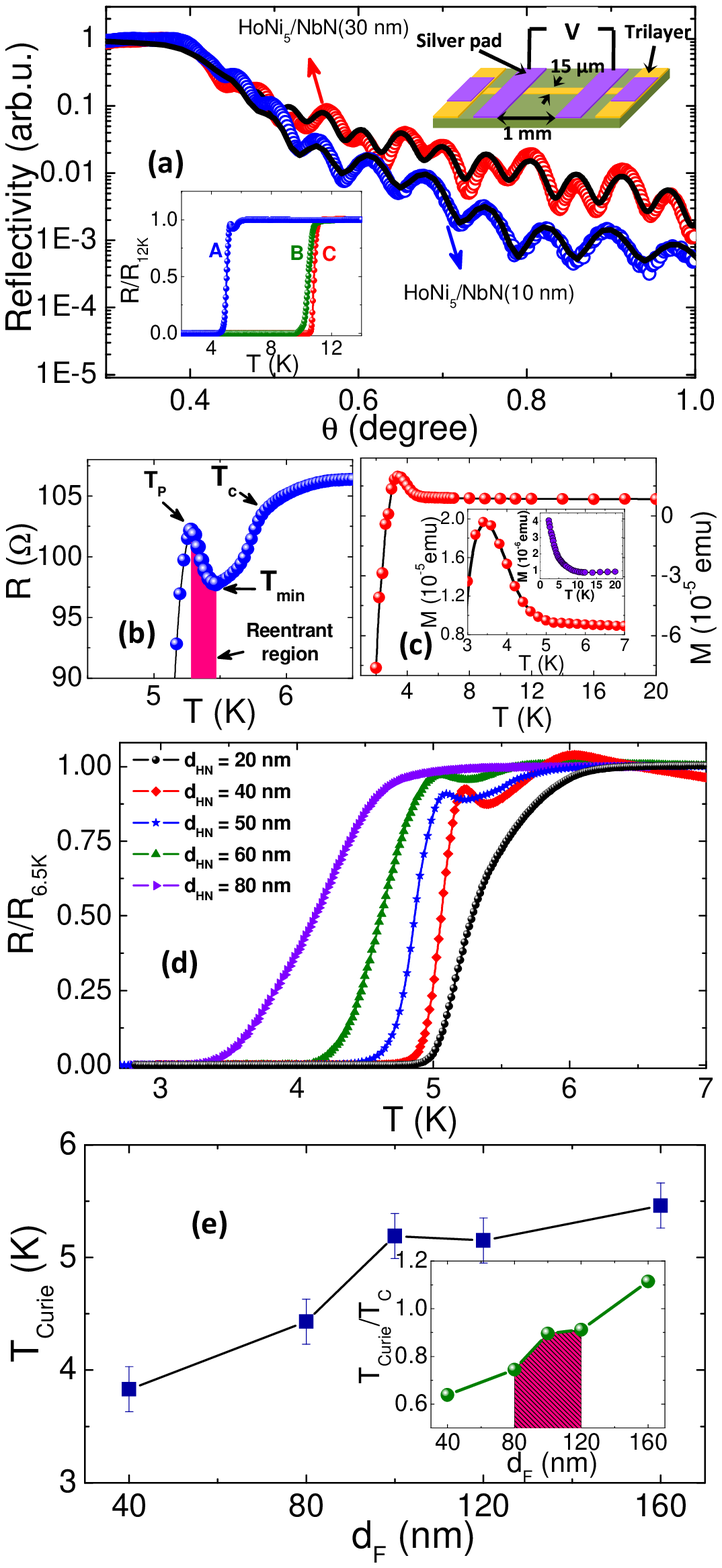}%
\end{center}
\caption{\label{fig1} (Color online) (a) X-ray reflectivity of HoNi$_5$/NbN bilayers shown as open circles with NbN thickness of 10 and 30 nm. Solid line is the best fitting of genetic algorithm. The upper inset shows the geometry of patterned sample for four probe measurements. The lower inset shows resistance vs. temperature of 10 nm thick NbN single layer (B) and HoNi$_5$/NbN/HoNi$_5$ trilayers with NbN thickness 10 nm (A) and 30 nm (C). (b) Larger view of R(T) for trilayer with 10 nm NbN. The thickness of each HoNi$_5$ layer in (b) and (c) is 50 nm. A colored band between 5.2 to 5.46 K indicates the region of reentrant superconductivity. (c) Magnetization vs. temperature of trilayer with 10 nm thick NbN, measured in 100 Oe out-of-plane field after zero field cooling of the film. The inset shows M vs. T between 3 to 7 K to highlight the magnetic ordering temperature of the film. It also shows M vs T for single layer HoNi$_5$ thin film measured after ZFC and applying the 100 Oe field in out-of-plane direction. A small but distinct ferromagnetic contribution to M(T) is seen at T $>$ T$_{Curie}$ which we believe comes from uncorrelated 4f spins of Ho. (d) R(T) of HoNi$_5$/NbN/HoNi$_5$ trilayers at zero field with five different thicknesses of HoNi$_5$ $\sim$20, 40, 50, 60, 80 nm. The NbN thickness was kept constant to $\sim$10 nm for all the samples. (e) Variation of the magnetic ordering temperature of the trilayer as a function of the total thickness of HoNi$_5$ layer. Inset show the dependence of T$_{Curie}$/T$_C$ on d$_F$. The shaded area range where reentrance is seen.}
\end{figure}

\begin{figure}[h]
\begin{center}
\includegraphics [width=8cm]{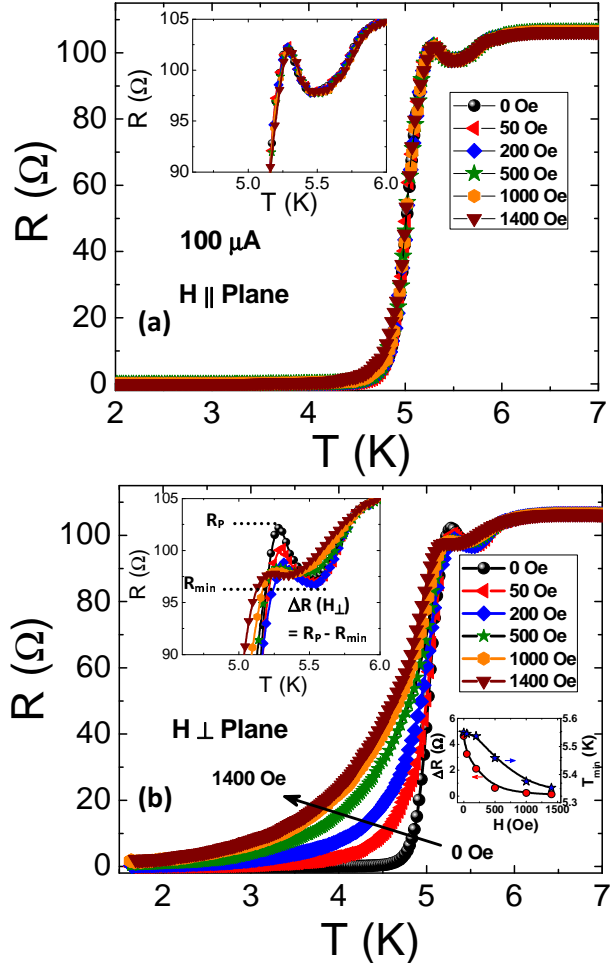}%
\end{center}
\caption{\label{fig2} (Color online) (a) Temperature dependent resistance at constant in-plane magnetic field varying from 0 to 1400 Oe. The inset shows larger view of the data for reentrant region. (b) R(T) with field applied along out-of-plane direction. A 100 $\mu$A current is applied along the bar for these measurements. The upper inset shows the clear view of reentrant behavior as a function of H$_{\perp}$. The peak and dip in R(T) is marked by R$_P$ and R$_{min}$ respectively. The lower inset shows the calculated value of $\bigtriangleup$R (= R$_{p}$ - R$_{min}$) and T$_{min}$ as a function of H$_{\perp}$}
\end{figure}

\begin{figure}[h]
\begin{center}
\includegraphics [width=8cm]{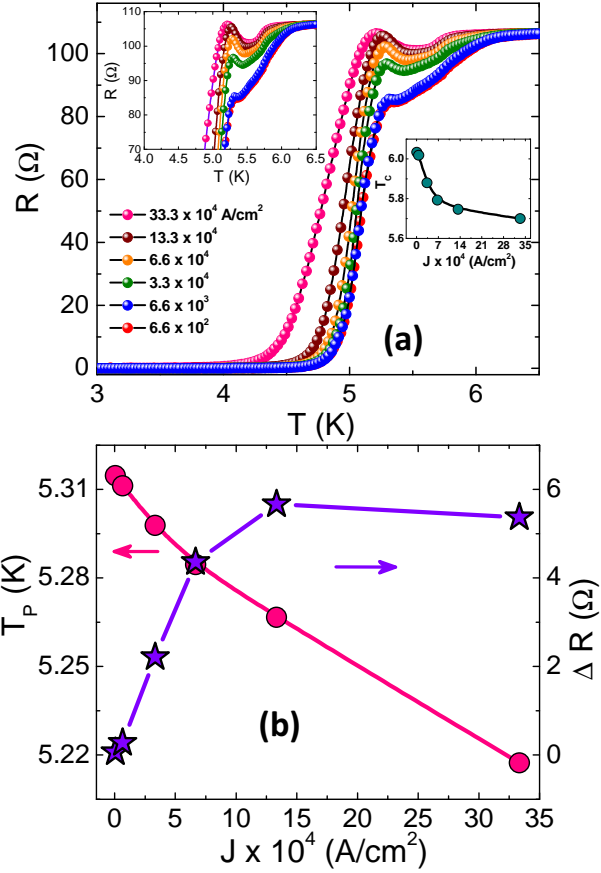}%
\end{center}
\caption{\label{fig3} (Color online) (a) Temperature dependent resistance measured in zero magnetic field with constant current density j between 6.6 x 10$^2$ A/cm$^2$ to 33.3 x 10$^4$ A/cm$^2$ (current 1 $\mu$A to 500 $\mu$A). The upper inset shows magnified view of R(T) where normal state reentrant behavior is observed. The behavior of critical temperature (T$_C$) with applied current is plotted in the lower inset. (b) Peak position (T$_P$) and $\Delta$R (= R$_{p}$ - R$_{min}$) as a function of current density calculated from panel (a).}
\end{figure}

\begin{figure}[h]
\begin{center}
\includegraphics [width=12cm]{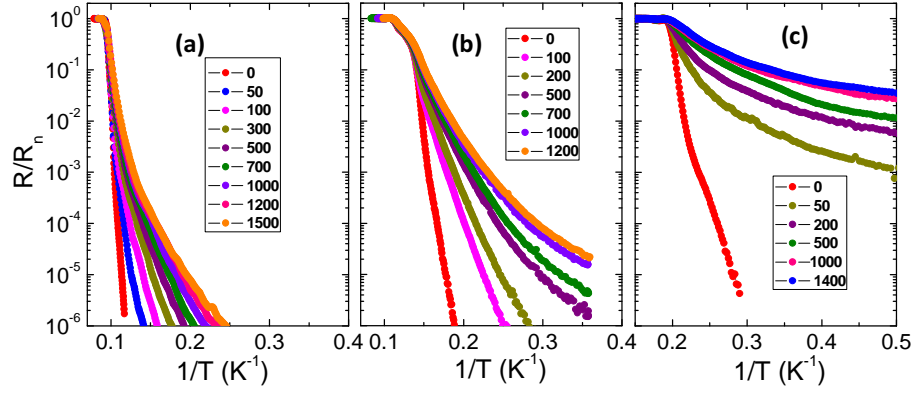}%
\end{center}
\caption{\label{fig4} (Color online) Normalized resistance vs. 1/T (Arrhenius plots) for (a) NbN(10 nm)/MgO (b) NbN(10 nm)/HoNi$_5$ (50 nm)/MgO bilayer and (c) HoNi$_5$ (50 nm)/NbN(10 nm)/HoNi$_5$ (50 nm)/MgO trilayer at various value of a magnetic field applied in the out-of-plane geometry (H$_{\perp}$).}
\end{figure}

\begin{figure}[h]
\begin{center}
\includegraphics [width=8cm]{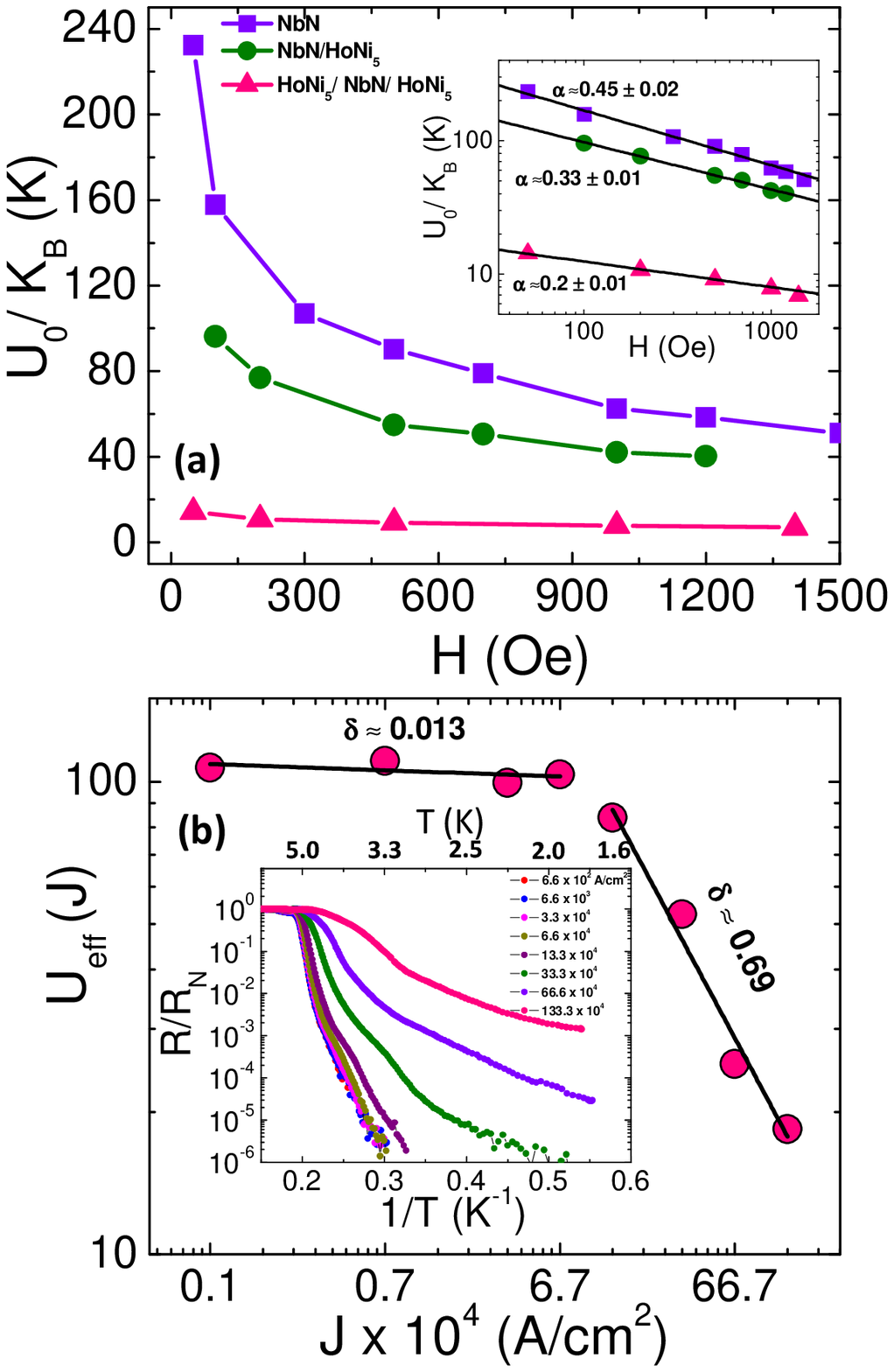}%
\end{center}
\caption{\label{fig5} (Color online) (a) Activation energy as a function of magnetic field applied perpendicular to plane of NbN(10 nm)/MgO, NbN(10 nm)/HoNi$_5$ (50 nm)/MgO bilayer and HoNi$_5$ (50 nm)/NbN(10 nm)/HoNi$_5$ (50 nm)/MgO trilayer. The inset shows plot in log-log scale. Solid line is the best fit to $U(H) \sim {H^{ - \alpha }}$. (b) Current dependence effective activation energy of the trilayer extracted from the Arrhenius plots shown in the inset of the figure. The value of $\delta$ is calculated from the best fitting of U$_{eff}$ $\alpha$ j$^{ - \delta }$. }
\end{figure}


\begin{thebibliography}{100}

\bibitem{Buzdin}
A. I. Buzdin, Rev. Mod. Phys. \textbf{77}, 935 (2005).

\bibitem{Lyuksyutov}
I. F. Lyuksyutov and V. L. Pokrovsky, Adv. Phys. \textbf{54}, 67 (2005).

\bibitem{Demler}
E. A. Demler, G. B. Arnold, and M. R. Beasley, Phys. Rev. B \textbf{55}, 15174 (1997).

\bibitem{Muhge}
Th. Muhge, K. Theis-Brohl, K. Westerholt, H. Zabel, N. N. Garifyanov, Yu. V. Goryunov, I. A. Garifullin, and G. G. Khaliullin, Phys. Rev. B \textbf{57}, 5071 (1998).

\bibitem{Cirillo}
C. Cirillo, S. L. Prischepa, M. Salvato, C. Attanasio, M. Hesselberth, and J. Aarts, Phys. Rev. B \textbf{72}, 144511 (2005).

\bibitem{Zdravkov}

V. Zdravkov, A. Sidorenko, G. Obermeier, S. Gsell, M. Schreck, C. Muller, S. Horn, R. Tidecks, and L. R. Tagirov
Phys. Rev. Lett. \textbf{97}, 057004 (2006).

\bibitem{Kehrle}
V. I. Zdravkov, J. Kehrle, G. Obermeier, S. Gsell, M. Schreck, C. Muller, H.-A. Krug von Nidda, J. Lindner, J. Moosburger-Will, E. Nold, R. Morari, V. V. Ryazanov, A. S. Sidorenko, S. Horn, R. Tidecks, and L. R. Tagirov
Phys. Rev. B \textbf{82}, 054517 (2010).

\bibitem{Garifullin}
I. A. Garifullin, D. A. Tikhonov, N. N. Garifyanov, L. Lazar, Yu. V. Goryunov, S. Ya. Khlebnikov, L. R. Tagirov, K. Westerholt, and H. Zabel, Phys. Rev. B \textbf{66}, 020505(R) (2002).


\bibitem{Tikhonov}
I. A. Garifullin, D. A. Tikhonov, N. N. Garifyanov, M. Z. Fattakhov, L. R. Tagirov, K. Theis-Brohl, K. Westerholt, and H. Zabel, Phys. Rev. B \textbf{70}, 054505 (2004).

\bibitem{Obi}
Y. Obi, M. Ikebe, and H. Fujishiro, Phys. Rev. Lett. \textbf{94}, 057008 (2005).

\bibitem{Verbanck}
G. Verbanck, C. D. Potter, V. Metlushko, R. Schad, V. V. Moshchalkov, and Y. Bruynseraede, Phys. Rev. B \textbf{57}, 6029 (1998).

\bibitem{Jiang}
J. S. Jiang, D. Davidovic, Daniel H. Reich, and C. L. Chien, Phys. Rev. Lett. \textbf{74}, 314 (1995).

\bibitem{Goryunov}
Th. Muhge, N. N. Garifyanov, Yu. V. Goryunov, G. G. Khaliullin, L. R. Tagirov, K. Westerholt, I. A. Garifullin, and H. Zabel, Phys. Rev. Lett. \textbf{77}, 1857 (1996).

\bibitem{Mercaldo}
L. V. Mercaldo, C. Attanasio, C. Coccorese, L. Maritato, S. L. Prischepa, and M. Salvato, Phys. Rev. B \textbf{53}, 14040 (1996).

\bibitem{Kontos}
T. Kontos, M. Aprili, J. Lesueur, and X. Grison, Phys. Rev. Lett. \textbf{86}, 304 (2001).

\bibitem{Ryazanov}
V. V. Ryazanov, V. A. Oboznov, A. Yu. Rusanov, A. V. Veretennikov, A. A. Golubov, and J. Aarts, Phys. Rev. Lett. \textbf{86}, 2427 (2001).

\bibitem{Oboznov}
V. A. Oboznov, V. V. Bolginov, A. K. Feofanov, V. V. Ryazanov, and A. I. Buzdin, Phys. Rev. Lett. \textbf{96}, 197003 (2006).
\bibitem{Robinson}
J. W. A. Robinson, S. Piano, G. Burnell, C. Bell, and M. G. Blamire, Phys. Rev. Lett. \textbf{97}, 177003 (2006); Phys. Rev. B \textbf{76}, 094522 (2007).

\bibitem{Khaire}
T. S. Khaire, M. A. Khasawneh, W. P. Pratt Jr., and N. O. Birge, Phys. Rev. Lett. \textbf{104}, 137002 (2010).

\bibitem{Witt}
J. W. A. Robinson, J. D. S. Witt, and M. G. Blamire, Science \textbf{329}, 59 (2010).

\bibitem{Bergeret}
F. S. Bergeret, A. F. Volkov, and K. B. Efetov, Rev. Mod. Phys. 77, 1321 (2005).

\bibitem{Leksin}
P. V. Leksin, N. N. Garifyanov, I. A. Garifullin, Ya. V. Fominov, J. Schumann, Y. Krupskaya, V. Kataev, O. G. Schmidt, and B. Buchner, Phys. Rev. Lett. 109, 057005 (2012)

\bibitem{Zhu}
J. Zhu, I. N. Krivorotov, K. Halterman, and O. T. Valls, Phys. Rev. Lett. 105, 207002 (2010).

\bibitem{Joshi}
Gyanendra Singh, P. C. Joshi and R. C. Budhani, Physica C \textbf{472}, 44 (2012).

\bibitem{Chien}
Chien-Te Wu, Oriol T. Valls and Klaus Halterman, Phys. Rev. Lett. \textbf{108}, 117005 (2012).

\bibitem{Eisaki}
H. Eisaki, H. Takagi, R. J. Cava, B. Batlogg, J. J. Krajewski, W. F. Peck, Jr., K. Mizuhashi, J. O. Lee, and S. Uchida, Phys. Rev. B \textbf{50}, 647 (1994).

\bibitem{Rathnayaka}
K. D. D. Rathnayaka, D. G. Naugle, B. K. Cho, and P. C. Canfield, Phys. Rev. B \textbf{53}, 5688 (1996).

\bibitem{Gupta}
L. C. Gupta, Advances In Physics \textbf{55}, 691 (2006).

\bibitem{Fertig}
W. A. Fertig, D. C. Johnston, L. E. DeLong, R. W. McCallum, M. B. Maple, and B. T. Matthias, Phys. Rev. Lett. \textbf{38}, 987 (1977).

\bibitem{Crabtree}
G. W. Crabtree, F. Behroozi, S. A. Campbell, and D. G. Hinks, Phys. Rev. Lett. \textbf{49}, 1342 (1982).


\bibitem{Gyanendra}
Gyanendra Singh, P. C. Joshi and R. C. Budhani, J. Appl. Phys. \textbf{109}, 113915 (2011).

\bibitem{Senapati}
K. Senapati, N. K. Pandey, Rupali Nagar, and R. C. Budhani, Phys. Rev. B \textbf{74}, 104514 (2006).


\bibitem{Gignoux}
D. Gignoux, A. Nait-Saada, A. Perrier de la Bthie, J. Phys. Coll. 40 (C5) (1979) 188.

\bibitem{Anderson}
P. W. Anderson and Y. B. Kim, Rev. Mod. Phys. \textbf{36}, 39 (1964).

\bibitem{Kes}
P. H. Kes, J. Aarts, J. van der Beek, and J. A. Mydosh, Supercond. Sci. Technol. \textbf{1}, 242 (1989).

\bibitem{Hsu}
J. W. P. Hsu and A. Kapitulnik, Phys. Rev. B \textbf{45}, 4819 (1992).

\bibitem{Budhani}
K. Senapati and R. C. Budhani, Phys. Rev. B \textbf{70}, 174506 (2004).

\bibitem{White}
W. R. White, A. Kapitulnik, and M. R. Beasley, Phys. Rev. Lett. \textbf{70}, 670 (1993).

\bibitem{Palstra}
T. T. M. Palstra, B. Batlogg, L. F. Schneemeyer, and J. V. Waszczak, Phys. Rev. Lett. \textbf{61}, 1662 (1988).; T. T. Palstra, B. Batlogg, R. B. van Dover, L. F. Schneemeyer, and J. V. Waszczak, Phys. Rev. B \textbf{41}, 6621 (1990).

\bibitem{Kucera}
J. T. Kucera, T. P. Orlando, G. Virshup, and J. N. Eckstein, Phys. Rev. B \textbf{46}, 11004 (1992).

\bibitem{Brunner}
O. Brunner, L. Antognazza, J.-M. Triscone, L. Miéville, and O. Fischer, Phys. Rev. Lett. \textbf{67}, 1354 (1991).

\bibitem{Fogel}
N. Ya. Fogel, V. G. Cherkasova, O. A. Koretzkaya, and A. S. Sidorenko, Phys. Rev. B \textbf{55}, 85 (1997).

\bibitem{Abulafia}
Y. Abulafia, A. Shaulov, Y. Wolfus, R. Prozorov, L. Burlachkov, Y. Yeshurun, D. Majer, E. Zeldov, H. Wuhl, V. B. Geshkenbein, and V. M. Vinokur, Phys. Rev. Lett. \textbf{77}, 1596 (1996).

\bibitem{Gennes}
P. G. de Gennes, Rev. Mod. Phys. 36, 225 (1964).

\bibitem{Fominov}
Y. Fominov thesis, arXiv:cond-mat/0311359.

\bibitem{Aarts}
J. Aarts, J. M. E. Geers, E. Bruck, A. A. Golubov, and R. Coehoorn, Phys. Rev. B \textbf{56}, 2779 (1997).

\bibitem{Sankar}
S. G. Sankar, D. A. Keller, R. S. Craig, W. E. Wallace, And V. U. S. RAO, J. Solid State Chemistry \textbf{9},78 (1974).

\bibitem{Bechman}
C. A. Bechman, W. E. Wallace and R. S. Craig J. Phs. Chem. Solids. \textbf{3} 463 (1974).

\bibitem{Lee}
P. A. Lee and T. V. Ramakrishnan, Rev. Mod. Phys. \textbf{57}, 287 (1985).

\bibitem{Malik}
S. K. Malik, F. J. Arlinghaus and W. E. Wallace, Phys. Rev. B \textbf{25}, 6488 (1982).

\bibitem{Leksin}
P. V. Leksin, N. N. Garifyanov, I. A. Garifullin, J. Schumann, V. Kataev, O. G. Schmidt, and B. Buchner, Phys. Rev. Lett. \textbf{106}, 067005 (2011).





\end{thebibliography}
\end{document}